\documentclass[amssymb,prl,preprint,aps]{revtex4}
\usepackage{graphicx}
\begin{document}
\title{Kondo physics in artificial molecules }
\author{K.Kikoin$^1$}
\affiliation{Department of Physics, Ben-Gurion University, Beer
Sheva 84105, Israel}
\author{Y. Avishai$^{1,2}$}
\affiliation{$^1$Department of Physics, Ben-Gurion University, Beer
Sheva 84105, Israel \\
$^2$ Department of Applied Physics, University of Tokyo}
\begin{abstract}
 Recent advancement in fabrication technologies enable the construction of
nano-objects with rather rich internal structures such as double
or triple quantum dots, which can then be regarded as artificial
molecules. The main new ingredient in the study of the Kondo
effect in such artificial (and also in natural) molecules is the
internal symmetry of the nano-object, which proves to play a
crucial role in the construction of the effective exchange
Hamiltonian. This internal symmetry combines continuous spin
symmetry $SU(2)$ and discrete point symmetry (such as mirror
reflections for double dots or discrete $C_{3v}$ rotation for
equilateral triangular dots. When these artificial molecules are
attached to metallic leads, the set of dot operators appearing in
the effective exchange Hamiltonian generate a group which is
refereed to as the dynamical symmetry group of the system
dot-leads [mostly $SO(n)$ or $SU(n)$], and the pertinent group
parameters (such as the value of $n$) can be controlled by
experiment. In this short review we clarify and expand these
concepts and discuss some specific examples. In particular we
concentrate on the difference between the chain geometry and the
ring geometry. When a perpendicular magnetic field is applied in
the ring geometry, its gauge symmetry $U(1)$ is involved in the
interplay with the  spin and orbital dynamics of the dot.
\end{abstract}
\maketitle
\section{Introduction: Kondo mapping and beyond}

There are numerous models in the literature of condensed matter
theory, whose significance for achieving progress in our
understanding of Nature goes far beyond the original aim of
explaining specific experimental observations. One may mention in
this context the Bardeen-Cooper-Schrieffer's explanation of the
nature of electron pairing in superconductors, the Ginzburg-Landau
equation intended for describing critical fluctuations, the
concept of self-localization of excitations in a perfect crystal
formulated by Deigen, Pekar and Toyozawa and various other seminal
ideas.  The explanation offered by J. Kondo for the puzzling
shallow minimum in the  temperature dependent resistivity of
metals  doped by magnetic impurities \cite{Kondo} is one of the
most salient examples of this kind of this scenarios.
To explain it, consider first Kondo original idea, which
was formulated within the framework of a well established
Hamiltonian describing exchange interaction between an impurity spin ${\bf
S}_{\bf r}$ located on a given site ${\bf r}$ and the spin density ${\bf s_{r}}$ pertaining to
a Fermi sea of conduction electrons at this site.
The latter is defined by the Fourier transform of the
itinerant spin ${\bf s}_{\bf kk'}= c^\dag_{\bf k\sigma}\hat \tau
c^{}_{\bf k'\sigma'}$ projected on the impurity site ${\bf r}$,
namely ${\bf s}_{\bf r}=\sum_{\bf kk'}{\bf s}_{\bf kk'}\exp{{\rm
i}({\bf k-k'})\cdot{\bf r}}$. Here $\hat \tau$ is the vector of
Pauli matrices for a spin 1/2. The so called $sd$-exchange
Hamiltonian is,
\begin{equation}\label{hex}
H_{\rm sd}=\sum_{\bf k,\sigma} \varepsilon_k c^\dag_{\bf k\sigma}
c^{}_{\bf k\sigma}+ J{\bf S}_{\bf r}\cdot {\bf s}_{\bf r}
\end{equation}
where $\varepsilon_k$ is the energy dispersion of the itinerant
electrons and $J$ is the exchange coupling constant. At first
glance, it looks deceptively simple. However, Kondo noticed that
 the first correction to the impurity scattering amplitude
of the conduction electrons beyond the Born approximation suffers
an infrared logarithmic divergence in energy or temperature. This
results in a singular behavior of amplitudes for an
antiferromagnetic sign of the exchange coupling (that is, $J>0$),
 and renders perturbation theory inapplicable below
a certain energy scale known as the Kondo temperature. Nearly two
decades of incessant efforts to take this singularity properly
into account and to find the ground state of the system had
crowned with finding the exact solution both numerically (in a
framework of Numerical Renormalization Group, NRG) \cite{Wil} and
analytically (by using the Bethe ansatz) \cite{Wig,Andr}.

Soon after Kondo's breakthrough, theoreticians started to extend
this promising conceptual framework for other physical situations
and for more complex objects than simple localized moments. It was
recognized that the Kondo mechanism should work also in systems
exhibiting electron tunneling, where two metallic slabs are
separated by thin dielectric layer, which forms a tunnel barrier
for electrons moving from  one slab to the other. It was shown
\cite{App,And66} that the magnetic impurity located somewhere near
the tunneling layer plays the same role in tunnel conductance as
magnetic impurity immersed in a metal (and subject to exchange
interaction with Fermi sea electrons) does for impurity
resistance. It was also shown \cite{SW} that the Friedel-Anderson
model \cite{Fried,And61} for resonance scattering of conduction
electrons by the electrons occupying the $3d$ levels of transition
metal impurities can be mapped on the exchange Hamiltonian
(\ref{hex}) provided the strong Coulomb interaction in the $3D$
shell suppresses charge fluctuations on the impurity site. In the
next stage of development orbital degrees of freedom were
incorporated in the Kondo physics. The idea of this generalization
is based on the fact that the magnetic impurity being put in the
center of coordinates imposes its point symmetry on the otherwise
translationally invariant crystal, and the appropriate description
of scattered waves should exploit the formalism of partial wave
expansion (either in spherical waves \cite{Fried} or in cubic
harmonics \cite{Dy}). Based on this idea, the generalized
Schrieffer-Wolff model was proposed \cite{Coq1,Coq2}, where the
magnetic impurity is described as an effective $N$-component
moment, but the exchange scattering is not restricted by the usual
spin selection rule $\Delta m=0,\pm1$ for the projection $m$ of
this moment. Another version of this model allots the impurity and
band states both by spin and orbital (referred to as "color" in
the general case) index. In case of spin $s=1/2$ and $N$ colors
the symmetry of the impurity  is $SU(2N)$ \cite{Afma88,Resa89}.
Appearance of additional degrees of freedom in the Kondo
Hamiltonian might lead to the interesting scenario of overscreened
Kondo effect, which arises when the number of conduction electron
"colors" exceed that of impurity moment \cite{Nobla}. This effect
is characterized by the non-Fermi-liquid low-temperature
thermodynamics unlike the standard Kondo effect, which only
modifies (although radically)) the Fermi liquid properties of
undoped metal \cite{Noz74}.

Another direction of expanding the Kondo physics is realized in
mapping the Kondo or Anderson Hamiltonian on those of other (not
necessarily magnetic) systems. A great variety of such
generalizations is possible because some quantum systems may be
described by a pseudospin, provided their low-energy states are
characterized as an effective two-level system (TLS) and external
perturbations allow transitions between these levels. The first
example of such generalization was suggested by the Hamiltonian
describing tunneling between a Fermi sea electrons and an atom
sitting in a double-well potential \cite{Zawad80}. Another
possibility of this sort arises when the crystal field splitting
is involved in formation of the low-energy states of impurity atom
(quadrupolar Kondo effect) \cite{Cox,Barnes}. This type of
"exotic" Kondo system was surveyed  in a detailed review
\cite{Coza}. One should mention also the possibility of involving
orbital degrees of freedom in the formation of Kondo-resonance for
an adsorbed transition metal impurity, where the orbital
degeneracy is lifted by the surface effects, e.g. the potential of
atomic step edge \cite{Zikl}.

A powerful incentive for further extension of the realm of Kondo
physics has been offered in 1988, when the idea of underbarrier
tunneling in presence of Kondo center was extended on the
tunneling between metallic electrodes and nanoobjects like quantum
dots or small metallic grains \cite{GR,Ng}. Such nanoobject  may
serve as a
 Kondo center (a localized moment)
 provided (i) the electron spectrum is discrete due
to spatial quantization, so that the level spacing
$\delta\epsilon$ exceeds the tunneling rate, (ii) Coulomb blockade
prevents charge fluctuations and (iii) the electron occupation
number is odd, so that the effective spin of the nanoobject is
1/2. In this case the tunneling Hamiltonian may be mapped on the
effective spin Hamiltonian (\ref{hex}), and the Kondo-like
singularity arises as a zero-bias anomaly (ZBA) in tunnel
transparency. This theoretical prediction was confirmed ten years
later \cite{Golgor,Cron,Sim} in the experiments on planar quantum
dots. Many experimental and theoretical studies then followed this
experimental breakthrough ever since.

In the course of developing this new realm of condensed matter
physics, further possibilities of extending the Kondo physics were
subsequently discovered and exploited. Original idea of using the
charge fluctuations as a source of Kondo tunneling was proposed
\cite{Matv} in the interim between the theoretical prediction of
Kondo tunneling and its experimental verification. It was shown in
this paper that in a situation, where two charge states of the
quantum dot with occupation $N,N+1$ are nearly degenerate, this
dot behaves as a two-level system, where the fluctuating charge
configuration plays part of pseudospin, whereas the real spin
projections may be treated as channel indices. Later on this idea
was further developed and modified \cite{Shilleb,Bol}.

Another facet of Kondo physics in nanoobjects was unveiled, when
the possibility of Kondo effect in quantum dots with even electron
occupation number was considered in several theoretical
publications \cite{KA00,PAK,Tag,Eto00}. In this case the quantum
dot with a singlet ground state may become magnetically active due
to external forces, and Kondo effect arises either at finite
energy \cite{KA00} or at finite magnetic field
\cite{PAK,Tag,Eto00}. Later on it was recognized that in many
cases the direct mapping of the original Kondo model into such
system is impossible, because the {\textsl{effective} symmetry of
the pertinent nanoobject is neither $SU(2)$ nor $SU(2N)$. The aim
of this review is to describe various physical situations where
the underlying nanoobject possesses complex (and in some sense
unusual) symmetries which are characterized by non-compact Lie
groups or combinations of such groups with discrete groups of
finite rotations.

\section{Surplus symmetries}

Among the sources of surplus symmetries which enrich Kondo physics
of nanoobjects one may find both discrete and continuous rotations
stemming from complicated geometrical configurations of complex
quantum dots, as well as those induced by external fields used in
practical realizations of nanodevices. In this short review we
will refrain from description of great variety of these devices,
which may be found in current literature (see e.g.,
\cite{Grab92,Kou97,Wiel03}). Fortunately, most of the relevant
physics may be exposed in a relatively simple situation
of electron tunneling through multivalley quantum dots in
contact with metallic electrodes (leads).

A multivalley dot is an island with electrons confined by
electrostatic potential in such a way that the spatially quantized
electrons are distributed between several valleys. These valleys
are coupled with each other by capacitive interaction and
tunneling channels. Up to now there are several realizations of
quantum dots with two and three valleys (double quantum dot, DQD,
and triple quantum dot, TQD, respectively). Experimentally, the
first such realizations of DQDs go back to mid 90-es
\cite{Hoff95,Mol95, Liver}. It was pointed out that these objects
can be treated as some forms of artificial molecules with each
constituent dot playing the role of an artificial atom
\cite{Palac,LodV,petpar,KA01}.

Compared with DQD, fabrication of TQD is a much more difficult
experimental task and the first experimental realizations of these
nanoobjects appeared only recently. One may mention in this
connection the realization of TQD with an "open" central valley
\cite{Marcus}. The term "open" here means that the tunneling
between the side dots and the adjacent leads is limited by strong
Coulomb blockade (see below), whereas the central dot freely
donates and accepts electrons to and from its own reservoir, so
its role in the device is only to mediate indirect exchange
between the two side dots. Another successful attempt to fabricate
a TQD was recorded in response to a theoretical proposal
\cite{Stopa} to use this geometry for realization of ratchet
effect in tunneling through nanoobjects. In this realization
\cite{Vid} the charge fluctuations were suppressed by the Coulomb
blockade mechanism in all three valleys. The feasibility of
filling the TQD with 1, 2 and 3 electrons by changing the gate
voltages was demonstrated quite recently \cite{Stud}.

Theoretical studies of electronics in the TQD geometry were also
inspired by possible applications in the field of quantum
information \cite{Tanam,Sarag}. Investigation of the Kondo physics
in TQD \cite{Brat} was motivated by the experimental observation
of molecular trimers $\rm{Cr}_3$ on gold sublayers \cite{Jam} by
means of the tunnel electron spectroscopy, which allows to observe
Kondo-type ZBA in conductance. Later on, other properties of these
trimers such as the two-channel Kondo effect \cite{Lazar,Affleck}
and the interplay between the Kondo effect and Aharonov-Bohm
effect in tunnel spectroscopy \cite{KKA06a} were considered
theoretically.

%\subsection{Devices with Double Quantum Dots}
Our main focus of interest is a theoretical modeling of a device consisting of a
multivalley quantum dot, metallic electrodes and corresponding gates. The
latter regulate the electron occupation of any particular valley and
of the dot as a whole. These properties and others are predetermined by the
geometrical position of the valleys relative to the metallic leads
(source and drain) in the device. Basically, there are three types of such devices which are
possible for describing electric circuits with
a DQD connected with a source  \textit{s} and drain \textit{d}.
They may be refereed  as sequential, parallel and T-shape connections (Fig.
\ref{config2}a,b,c, respectively). The two small dots
which combine to form the DQD may either be
identical or may differ in their size. Besides, different gate
voltages may be applied to different valleys.
\begin{figure}[htb]
\centering

\includegraphics[width=65mm,angle=0]{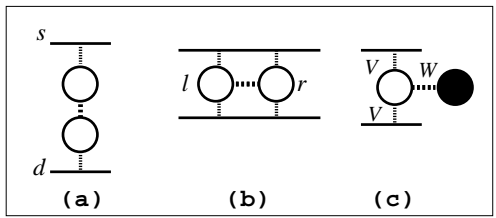}
\caption{Double quantum dot in sequential (a), parallel (b) and
T-shape (c) geometries. Filling black marks the valley detached
from the leads.} \label{config2}
\end{figure}

In the most symmetrical configuration of identical dots the only
additional symmetry which characterizes a DQD is the permutation
symmetry $P_2$. In close analogy with Kondo effect in magnetically
doped metals, one should consider the total symmetry of a device
'source + DQD + drain'. Then the symmetry group of the sequential
configuration (Fig. 1a) contains the only discrete element, namely
the \textit{s-d} reflection axis, and this element adds nothing to
the $P_2$ symmetry of the isolated DQD. The same statement is
valid for the T-shape geometry (Fig. \ref{config2}c). In the
parallel geometry the system as a whole possesses two reflection
axes, namely source-drain \textit{s-d} and left-right
\textit{l-r}, where the indices \textit{l} and \textit{r} label
two valleys of DQD.

Triple quantum dots (TQD) present theoreticians (and
experimentalists as well) with a richer variety of geometrical
configurations and possess more symmetries (Fig. \ref{config3}).
Similarly to DQD these trimers may be oriented both in sequential
(vertical) and parallel (lateral) geometries (Fig. \ref{config3}a
and \ref{config3}b, respectively). The natural generalization of
the T-shape geometry presented in Fig. 1c is the cross geometry
(Fig. \ref{config3}c). Besides, TQD may be organized in a form of
a triangle, which may form a closed or open element in an
electronic circuit (Figs. \ref{config3}d,e and \ref{config3}f,
respectively). In the two latter cases one deals with a
three-terminal tunnel device. We will call the conformations shown
in Figs. \ref{config3}e and \ref{config3}f ``ring'' and ``fork''
configurations.
\begin{figure}[htb]
\centering
\includegraphics[width=65mm,angle=0]{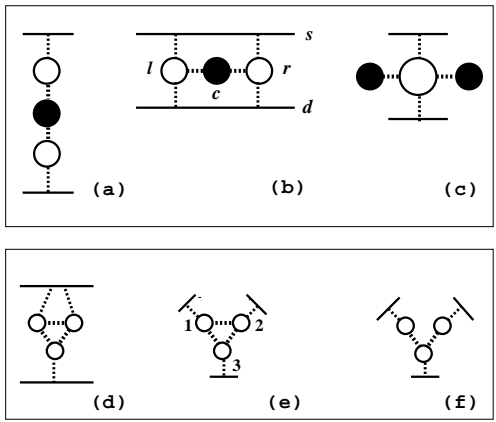}
\caption{Triple quantum dot in sequential (a), parallel (b),
cross-shape (c), two-terminal triangular (d), three-terminal
triangular (e) and fork (f) geometries.} \label{config3}
\end{figure}

Now let us discuss the discrete symmetry elements characteristic
for the above TQD configurations. If all three valleys are
equivalent, the discrete symmetry of an isolated linear TQD is
that of the permutation group $P_3$. The contact with the leads
adds one more symmetry element, the \textit{s-d} reflection,
provided all three dots are coupled with the leads. We however
consider the devices, where the central dot (filled black) is not
coupled directly with the leads. Then the $P_3$ symmetry is lost.
The only discrete symmetry element, namely \textit{s-d} reflection
is left in a vertical geometry (Fig. \ref{config3}a), whereas both
\textit{s-d} and \textit{l-r} reflections characterize a symmetry
of lateral TQD (Fig. \ref{config3}b). The same is valid for the
cross geometry of Fig. \ref{config3}c. The $P_3$ symmetry is
inherent in the three-terminal configuration of Fig.
\ref{config3}e. In this case it is better to use the
classification of discrete rotation group $C_{3v}$, which is
isomorphic to the permutation group $P_3$. One may say
\cite{KKA06a,KAKE} that in the perfect triangular configuration
the TQD imposes its $C_{3v}$ symmetry on the device as a whole in
close analogy with the Coqblin-Schrieffer-Cornut version
\cite{Coq1,Coq2} of the conventional Kondo problem. In the
geometries of Fig. \ref{config3}d,f the only element which is
remained of the original symmetry of triangle is the \textit{l-r}
symmetry like in the cross geometry. \textit{External magnetic
field} applied perpendicularly to the plane of the triangle lowers
the symmetry of a device by adding one more element, that is,
chirality of the electron tunneling from the source to the drain.

Before considering the consequences of these symmetries for the
Kondo physics, several introductory remarks about the derivation
of the Kondo Hamiltonian are in order. In an early period of the
theoretical studies, the problem of electron tunneling through
short chains of quantum dots under strong Coulomb blockade
restrictions was formulated in terms of the Mott-Hubbard picture
\cite{Stads,Klim}. The theory was based on the idea that electrons
injected from the source \textit{do not lose coherence} when
propagating through the sequence of quantum dots until they leave
the chain for the drain electrode. This approach is valid only for
short enough ``Hubbard chain'', where the tunneling $W$ between
the adjacent valleys exceeds the tunneling $V$ between the dot and
the metallic leads. Using generalized Landauer method, one
describes the tunnel transparency in terms of the Green functions
of a nanoobject in contact with the leads \cite{Mew}. Such a
procedure starts with diagonalization of the Hamiltonian of
nanoobject with subsequent calculation of renormalization of the
spectrum of quantum dot due to tunnel contact with the leads.
Early studies of these problems concentrated on the calculations
of the Coulomb blockade peaks which arise with changing the
occupation of the valleys (so called Coulomb staircases). In terms
of the Hubbard model, the Coulomb resonances are the Hubbard
"minibands" \cite{Stads}, which arise as a result of collective
Coulomb blockade \cite{Golhal} (Hubbard repulsion). It is known,
however, that the spectral function of the Hubbard model contains
also the central peak of predominantly spin origin. This peak is
responsible for zero-bias anomalies in the tunnel conductance,
which are at the center of our attention.

If the "chain" contains a single dot, one deals with the
conventional mapping of the Anderson-like tunneling problem onto
the Kondo-like cotunneling problem \cite{GR,Ng}, so that the
central peak is indeed the famous Abrikosov-Suhl resonance pinned
to the Fermi level of the band electrons \cite{Wig,Andr,Langr},
which is responsible for the ZBA in tunnel conductance. DQD in
sequential geometry is the first non-trivial generalization of the
single dot case, and one of the primary tasks is to look whether
some qualitative differences from the conventional Kondo effect
arise because of combining the features of Anderson and Hubbard
models in the effective tunneling Hamiltonian.

In any case, working in the above paradigm \cite{Stads,Klim}, one
should derive the effective exchange Hamiltonian $H_{\rm ex}$ in
accordance with the following procedure.  The starting Hamiltonian
is chosen in the same form as the original Anderson impurity
Hamiltonian \cite{And61}:
\begin{equation}\label{and2}
H = H_{\rm band}+ H_{\rm dot} + H_{\rm t}
\end{equation}
where $H_{\rm dot}$ describes the properties of a chain detached
from the lead \textit{in terms of its eigenstates}
$|\Lambda\rangle$:
\begin{equation}\label{dot2}
H_{\rm dot}=\sum_\Lambda E_\Lambda |\Lambda\rangle\langle\Lambda|
+ Q(\hat N -{\cal N}).
\end{equation}
Here $\hat N$ is the operator of total electron number in the dot.
The last term in (\ref{dot2}) describes the Coulomb blockade
mechanism: the total occupation of the dot $\cal N$ in neutral and
charged states is fixed by the Coulomb blockade parameter $Q$
entering the capacitive energy of a complex quantum dot as a
whole. The eigenvalues $E_\Lambda$ are found at fixed occupation
$\cal N$ of DQD or TQD with all tunneling matrix elements $W$ and
interdot capacitive interactions $Q'$ taken into account. The
tunneling Hamiltonian $H_{\rm t}$ intermixes the states  from
adjacent charge sectors ${\cal N}, \cal N'$ due to injection of
extraction of an electron from the complex dot:
\begin{equation}\label{tun2}
H_{\rm t}=\sum_{\Lambda\Lambda'}\sum_{b=s,d}\sum_{k\sigma}
V^{\Lambda\Lambda'}_{bk\sigma} c^\dag_{bk\sigma}
X^{\Lambda\Lambda'} + {\rm H.c}.
\end{equation}
where $X^{\Lambda\Lambda'}= |\Lambda\rangle\langle\Lambda'|$ are
the universal Hubbard operators \cite{Hub}. The tunnel parameters
$V^{\Lambda\Lambda'}_{bk\sigma}$ are usually approximated by a
single parameter $V$.  In the Hamiltonian (\ref{tun2}) these
configuration changing operators describe transitions between the
states belonging to different charge sectors (one of this sectors
corresponds to the neutral CQD and another belong to positively or
negatively charged CQD. The index $b$ enumerates the leads
($b=s,d$ in the two-terminal configurations)

The Hamiltonian $H_{\rm band}$ has the standard form
\begin{equation}
H_{\rm band}=\sum_{bk\sigma} \varepsilon_k c^\dag_{bk\sigma}
c^{}_{bk\sigma}.
\end{equation}
Comparing to the corresponding term in (\ref{hex}), this
Hamiltonian contains one more index $b$. In the geometries with
\textit{s-d} reflection symmetry one may rotate the frame in such
a way that the band operators are classified as even and odd
operators relative to this reflection
\begin{equation}\label{evod}
c_{ek\sigma}=2^{-1/2}(c_{sk\sigma}+c_{dk\sigma}),~~~
c_{ok\sigma}=2^{-1/2}(c_{sk\sigma}-c_{dk\sigma})
\end{equation}
In case of single and double quantum dots this rotation usually
excludes the odd combination from the tunneling Hamiltonian
\cite{GR}. However in case of TQD it is not necessarily the case.
We will return to this question in section IV.

The cotunneling (exchange) Hamiltonian is usually obtained from
(\ref{and2}) by means of the Schrieffer-Wolff (SW) canonical
transformation \cite{SW}, which excludes the states
$|\Lambda\rangle$ belonging to the charge sectors ${\cal N}\pm 1$
from the effective Fock space. At fixed ${\cal N}$ we are left
solely with spin degrees of freedom. In a conventional situation,
the relevant symmetry is $SU(2)$ and the SW procedure results in
an effective Hamiltonian (\ref{hex}). In case of even ${\cal N}$
and geometries including discrete rotations, the situation is more
complicated, and the SW procedure intermixes the states
$|\Lambda\rangle$ belonging to different irreducible
representations of the Hamiltonian (\ref{dot2}). The corresponding
terms in the effective Hamiltonians may be expressed by means of
the corresponding Hubbard operators $X^{\Lambda\Lambda'}$. In many
cases, combinations of these operators form closed algebras which
generate non-compact groups $SO(N)$ or $SU(N)$ with $N>2$,
describing the \textit{dynamical symmetry} of complex quantum
dots. Involvement of these dynamical symmetries turns
 the procedure of mapping the tunnel problem onto
an effective exchange problem  to be more
complicated than in the simpler situations which were briefly
described in Section I. New features of the Kondo effect arising as a
result of this procedure were described for the first time using the
 T-shaped DQD as an example \cite{KA01}. Various manifestations of
dynamical symmetries in physical problems are described in the
monographs \cite{Eng,Mama}. Some mathematical aspects of the
dynamical symmetries as applied to the Kondo problem may be found
in the recent reviews \cite{Nova,kisrew}.

\section{Kondo physics for short chains}

Short chains represented in Figs. \ref{config2}a-c,
\ref{config3}a-c are the most elementary objects, where many
aspects of Kondo tunneling beyond the original paradigm
\cite{App,GR} of Kondo mapping may be demonstrated. There is much
in common between the linear DQD and TQD in vertical and lateral
geometries, although there are some effects specific only form
T-shaped and cross-shaped configurations (Figs. \ref{config2}c and
\ref{config3}c), which will also be emphasized in this section.

\subsection{Double quantum dots}

We start a more detailed discussion of Kondo tunneling through
artificial molecules with the case of DQD in the vertical geometry
(Fig. \ref{config2}a). Historically, this is the first
generalization of a single dot problem (see above). At present,
the tunneling through this simplest DQD is well understood.
Following an extensive theoretical discussion of Kondo physics in
vertical QD geometry
\cite{Ivan,Pohj,Aono,GeMe,GoLo,Bord03,TaKa,SaKa}, it turned out
that the most general description of Kondo tunneling through
vertical DQD may be given in terms of $SU(4)$ and $SO(4)$
symmetries of a low-energy multiplets in cases of odd and even
electron occupation ${\cal N}$, respectively.

As was mentioned above, the  microscopic description of Kondo
tunneling is analyzed in a framework of the generalized Anderson
Hamiltonian (\ref{and2}). All generic features of Kondo mapping
are seen already in the most elementary cases of ${\cal N}=1,2$.
DQD in these charge sectors can be treated as an artificial analog
of the molecular ion ${\rm H}^+_2$ and the neutral molecule ${\rm
H}_2,$ respectively.

In case of ${\cal N}=1$ the eigenstates of the Hamiltonian $H_{\rm
dot}$ (\ref{dot2}) are
\begin{equation}\label{2.1}
 E_{1,2} = \epsilon_d \mp W~,
\end{equation}
 where $\epsilon_d$ is the discrete level position in the
isolated valley of the DQD and $W$ is the inter-valley tunneling. These
two levels forming the "Hubbard miniband" in the ${\cal N}=1$
charge sector correspond to even and odd combinations of the
electron wave functions in the double-well confinement potential
of the DQD. We are interested here not in the resonance tunneling
through these levels, but in the ZBA connected with the Kondo
effect. The characteristic energy $T_K$ which scales  the Kondo
effect, should be compared with the level splitting $W$. However
$T_K$ itself depends on the level splitting, so that the
comparison procedure should be performed self-consistently. In the
limiting case $T_K(W)\ll W$, one may ignore the odd state $E_2$.
Since the odd combination of the lead electron states in
(\ref{evod}) is also excluded from the problem, we immediately see
that in this case the mapping procedure (SW transformation)
reduces the tunneling problem to the case of single quantum dot
with the Hamiltonian (\ref{hex}), where the exchange constant $J$
is estimated as $J=V^2/E_C$, and $E_C$ is expressed via addition
and extraction energies, i.e. the energy costs to add or subtract
an electron on/from the quantum dot:
\begin{eqnarray}
&&E_C^{-1}= E_+^{-1}+E_-^{-1} \\
&&E^+=\epsilon_d+Q-\epsilon_F,~~~ E^-=\epsilon_F-\epsilon_d~.
\nonumber
\end{eqnarray}
The Kondo temperature is given by the standard equation
$T_K=D\exp(-1/2\rho J)$, where $D$ is the characteristic energy
scale for the band electrons in the leads and $\rho$ is the
density of states at the Fermi level $\epsilon_F$.

In the opposite limit $T_K(W)\gg W$, one may neglect $W$ when
calculating $T_K$, so that the dot level acquires "orbital"
degeneracy. Due to this quasi degeneracy the effective exchange
acquires an additional factor 2, so that the Kondo temperature in
this case is $T_K=D\exp(-1/4\rho J)$. If fact, the difference
between the two limiting cases is the manifestation of the $SU(4)$
symmetry, which characterizes the spin state of an electron in the
double well potential. This symmetry will be described more
strictly when we will turn to the case of lateral DQD. One should
note that here we encountered the first manifestation of the
complicated structure of spin multiplets in complex quantum dots,
namely with non-universality of $T_K$: it crucially depends not
only on the parameters of the Hamiltonian but also on the
effective symmetry of the low-lying spin states involved in Kondo
cotunneling. Interpolation between two limiting cases may be
described in terms of gradual $SU(2)\to SU(4)$ crossover.

The study of the charge sector ${\cal N}=2$ uncovers another
important aspect of the Kondo mapping procedure. This sector
corresponds to the half-filled Hubbard chain, where the
single-electron tunneling is suppressed by Coulomb blockade (the
interior of the Coulomb diamond in terms of the theory of
single-electron tunneling \cite{Grab92,Kou97}). On the surface of
it, the Kondo tunneling is also not achievable because the ground
state of the DQD with ${\cal N}=2$ is a spin singlet in close
analogy with the case of hydrogen molecule $H_2$. However, the
dynamical symmetry of DQD plays its part in this case as well.

Indeed, in case of strong Coulomb blockade $\beta\equiv W/Q \ll 1$
the spectrum of isolated DQD consists of  two low-lying spin
states $E_{S,T}$ and two charge transfer excitons $E_{e,o}$ (even
singlet and odd triplet), with
\begin{eqnarray}
E_S & = & 2\varepsilon-2\beta W,~~~ E_T  =  2\varepsilon
\label{2.2} \\
E_{o} & = & 2\varepsilon +Q,~~~ E_{e}  =  2\varepsilon +Q +2\beta
W~ . \nonumber
\end{eqnarray}
Only the low-energy Singlet/Triplet (S/T) pair is relevant for
Kondo tunneling. Like in the case of ${\cal N}=1$, the triplet
state $E_T$ is frozen provided the exchange gap $\Delta_{\rm ex}=
2\beta W$ essentially exceeds $T_K$. In the opposite case
$\Delta_{\rm ex}\ll T_K$ the spin multiplet as a whole is involved
in Kondo tunneling, and the Kondo-type ZBA may survive
\cite{GeMe}. One may describe this phenomenon in terms of the
theory of conventional two-site Kondo effect \cite{JoVa}.
According to this theory, the antiferromagnetic intersite exchange
$J_{12}$ competes with the single-dot Kondo temperature $T_{K0}$.
At small $J_{12}$ (small enough $W$ in our case) each spin is
screened independently and the Kondo-type ground state may be
achieved. However, there exists a critical value $J_c$, so that at
$J_{12}>J_c$ the two spins are locked into a singlet state and the
Kondo effect does not apply.

The relevant dynamical symmetry is the $SO(4)$ symmetry of S/T
manifold ("spin rotator" \cite{KA01}). Again we postpone the
derivation of the effective Hamiltonian for the Kondo tunneling
and the discussion of its observable manifestations for the case
of lateral DQD. Here one should note that in the charge sector
${\cal N}=2$ the exchange gap $\Delta$ plays the same role as the
charge-transfer gap W in case of ${\cal N}=1$: the dependence
 $T_K(\Delta)$ is determined by a gradual symmetry crossover.

Now we turn to the parallel (lateral) geometry of Fig.
\ref{config2}b. A new element, which arises in this case is the
possibility of \textit{separate} channel for each dot. The new
features brought by this additional quantum number were discussed
in several publications \cite{Eto05,HoSch}. In case of two
channels the conduction electrons retain additional "color" after
rotation (\ref{evod}). We will enumerate the states corresponding
to this color by the same indices $i,j=1,2$ as those used for the
dots because they describe the odd and even states relative to the
\textit{l-r} reflection like those in the states
(\ref{2.1}),(\ref{2.2}).

We start once more with the charge sector ${\cal N}=1$, where a
single electron in DQD is distributed between two wells. It is
well known that the electron in a double well may be described by
means of a pseudospin operator ${\bf T}$ with the following components"
\begin{equation}\label{pseudo}
T^z = \sum_\sigma \left( d^\dag_{2\sigma}d^{}_{2\sigma} -
d^\dag_{1\sigma}d^{}_{1\sigma}\right),~~~T^+ = \sum_\sigma
d^\dag_{2\sigma}d^{}_{1\sigma}~~~T^- = \sum_\sigma
d^\dag_{1\sigma}d^{}_{2\sigma}.
\end{equation}
This vector, together with four spin vectors ${\bf S}_{ij}$ with
the following components,
\begin{equation}
S^z_{ij} = \frac{1}{2}\left( d^\dag_{i\uparrow}d^{}_{j\uparrow} -
d^\dag_{i\downarrow}d^{}_{j\downarrow}\right),~~~S^+_{ij} =
d^\dag_{i\uparrow}d^{}_{j\downarrow}~~~S^-_{ij} =
d^\dag_{i\downarrow}d^{}_{j\uparrow}
\end{equation}
form the set of 15 generators for the $SU(4)$ group.

To close the basis for the effective spin Hamiltonian, one has to
introduce similar vector operators for the electrons in the leads,
namely, the pseudospin operator \textbf{t} with components
\begin{equation}
t^z = \sum_{kk',\sigma} \left( c^\dag_{2k\sigma}c^{}_{2k'\sigma} -
c^\dag_{1k\sigma}c^{}_{1k'\sigma}\right),~~~t^+ =
\sum_{kk',\sigma} c^\dag_{2k\sigma}c^{}_{1k'\sigma}~~~t^- =
\sum_{kk',\sigma} c^\dag_{1k\sigma}c^{}_{2k'\sigma}.
\end{equation}
and four spin operators ${s}_{ij}$ with components
\begin{equation}
s^z_{ij} = \sum_{kk'} \left( c^\dag_{ik\uparrow}c^{}_{jk'\uparrow}
- c^\dag_{ik\downarrow}c^{}_{jk'\downarrow}\right),~~~s^+_{ij} =
\sum_{kk'} c^\dag_{ik\uparrow}c^{}_{jk'\downarrow}~~~s^-_{ij} =
\sum_{kk'} c^\dag_{ik\downarrow}c^{}_{jk'\uparrow}.
\end{equation}

The SW transformation carried out in terms of these operators results in
the effective Hamiltonian,
\begin{equation}\label{xeff1}
H_{\rm eff}= H_{\rm band} +H_{\rm dot} + 2\sum_{ij}J_{ij}{{\bf
S}_{ij}}\cdot {{\bf s}_{ji}} + 2K {\bf T}\cdot {\bf t}~.
\end{equation}
In the fully symmetric case which we describe here, all the
effective exchange constants have the same value $J_{ij}=K\equiv
J.$ Then the Hamiltonian (\ref{xeff1}) may be reduced to a more
compact and familiar form of exchange Hamiltonian in a fictitious
magnetic field \cite{Eto05}
\begin{equation}\label{xeff2}
H_{eff}=H_{\rm band}+ J{\bf S}\cdot{\bf s} - \widetilde{\bf
B}\cdot{\bf s}~,
\end{equation}
where
\begin{eqnarray}
{\bf S}=\frac{1}{2}\sum_{\alpha\beta}f^\dag_\alpha \hat
\Sigma_{\alpha\beta} f^{}_\beta,~~~{\bf
s}=\frac{1}{2}\sum_{kk'}\sum_{\alpha\beta}c^\dag_\alpha \hat
\Sigma_{\alpha\beta} c^{}_\beta~.
\end{eqnarray}
Spin fermion operators $f^{}_\alpha$  and the matrix $\hat \Sigma$
involve 15 components, which are generators of the Lie algebra
\textsl{su}(4), determined as
$$
\{(\tau^+,\tau^-,\tau^z,I)\otimes{\sigma^+,\sigma^-,\sigma^z,I}\}-\{I\otimes
I\}
$$
where $\sigma^\nu (\tau^\nu)$ are the Pauli matrices in the spin
(pseudospin) space and $I$ is the unit matrix. The fictitious
magnetic field has only one non-zero component, namely the
$\tau_z\otimes I$ component, and its magnitude is  $W$, so that
$\widetilde{\bf B}\cdot{\bf s}=WT^z$. Thus one concludes that in
DQD with odd occupation the formally multichannel Kondo
Hamiltonian (\ref{xeff1}) may be mapped onto the standard
Hamiltonian (\ref{xeff2}) in the case of complete channel
degeneracy, and the contribution of additional (permutation)
degrees of freedom is described in this degenerate case by a
fictitious magnetic field

In the charge sector ${\cal N}=2$ the permutation symmetry
degenerates into trivial unit transformation. As to the spin
degrees of freedom, one may try to described them by means of two
spin operators ${\bf s}_l$ and ${\bf s}_r$ using the above
mentioned analogy with the two-site Kondo problem \cite{JoVa}.
However, such approach \cite{PG01} should be used with some
caution. One should take into account that in a situation, where
both triplet and singlet two-electron states are involved in Kondo
effect, these spins are non-independent because the kinematical
constraint is imposed on the S/T manifold by the Casimir operator
${\cal C} \neq {\bf s}_l^2+{\bf s}_r^2$. In accordance with the
prescriptions of the theory of dynamical symmetries
\cite{Eng,Mama,Nova}, one should construct two operators
\begin{equation}
{\bf S}={\bf s}_l+{\bf s}_r, ~~~ {\bf R}={\bf s}_l-{\bf s}_r
\end{equation}
and impose on them the kinematic constraint
\begin{equation}
{\cal C} = {\bf S}^2+{\bf R}^2=3
\end{equation}
Then the three components $(S^z,S^+, S^-)$ of the vector ${\bf S}$
describe  the states within the spin triplet and transitions
between them, whereas the three components $(R^z,R^+, R^-)$ of the
vector ${\bf R}$ describe transitions between the singlet S and
the states with spin projections $\mu=1,0,-1$ of the triplet T.
Six components of the vectors ${\bf S}$,${\bf R}$ form a closed
algebra
\begin{eqnarray} &&[S_\alpha,S_\beta]  =
ie_{\alpha\beta\gamma}S_\gamma,~[R_\alpha,R_\beta]=
ie_{\alpha\beta\gamma}S_\gamma,~
[R_\alpha,S_\beta]=ie_{\alpha\beta\gamma}R_\gamma . \label{comm1}
\end{eqnarray}
and form a set of generators of $SO(4)$ group. Here
$\alpha,\beta,\gamma$ are Cartesian coordinate indices, and
$e_{\alpha\beta\gamma}$ is the anti-symmetric Levi-Civita tensor.
Two vector operators are orthogonal, ${\bf S\cdot R} = 0.$ Under
these constraints, two vectors ${\bf s}_{l,r}$ are rather
fictitious than real spin operators. More detailed discussion of
interconnections between two representations as well as the
derivation of these operators by means of the Hubbard operators
$X^{\Lambda\Lambda'}$ may be found in \cite{KA01}. If one
tunneling channel couples this ``spin rotator" with the reservoir
of conduction electrons, then the dynamical group $SO(4)$ exhausts
the spin degrees of freedom involved in Kondo tunneling. This
scenario is realized in the  T-shape geometry of Fig.
\ref{config2}c (see below).

More complicated is the mapping procedure for the two-channel
Anderson Hamiltonian describing the DQD shown in Fig
\ref{config2}b. In this case the discrete symmetry is explicitly
involved in the cotunneling process, so that the SW transformation
 give the exchange part of $H_{\rm
eff}$ in the form
\begin{equation}\label{xeff3}
H_{\rm ex}= 2\sum_{ij}J_{ij}{{\bf S}_{ij}}\cdot {{\bf s}_{ji}} +
2\sum_{ij}\tilde J_{ij}{{\bf R}_{ij}}\cdot {{\bf s}_{ji}} + 2K
{\bf T}\cdot {\bf t}~.
\end{equation}
DQD with \textit{l-r} reflection axis, where all the states are
classified as even or odd states relative to this reflection, is
described by this Hamiltonian may be qualified as a ``double spin
rotator'' \cite{KKA04}. The corresponding dynamical symmetry group
is $P_2\times SO(4)\times SO(4)$

Before turning to the physical aspects of Kondo tunneling through
the objects of $SU(4)$ and $SO(4)$ symmetries, one should mention
that the DQD in lateral geometry with two channels is described by
the same basic two-level Anderson Hamiltonian as the single planar
quantum dot with two levels, one of which is occupied and another
is empty. The latter model was considered in many publications
\cite{PAK,Tag,Eto00,HoSch,ISS,PG01,PHG,HoZa} (see also the review
papers \cite{Eto05,PG04}). So the physical manifestations of
"variable" symmetry are common for both systems. Fig. \ref{fig3}
illustrates the variation of $T_K$ as a function of control
parameter (the tunnel splitting $W$ in case of ${\cal N}=1$ and
the exchange gap $\Delta_{\rm ex}$ in case of ${\cal N}=2$). In
both cases the maximum value of $T_K$ correspond to the degeneracy
points.
\begin{figure}[htb]
\centering
\includegraphics[width=75mm,angle=0]{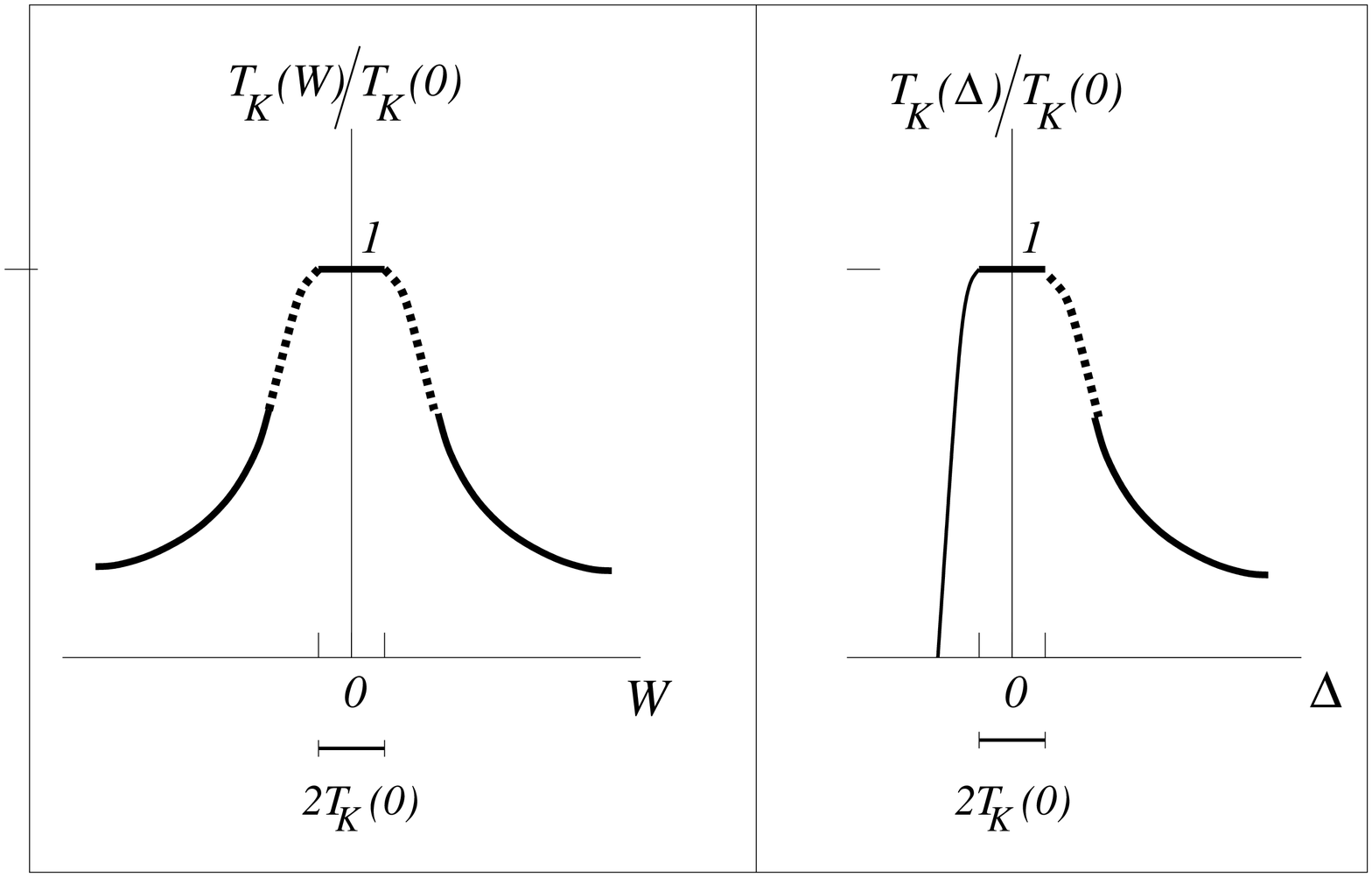}
\caption{Dependence of the Kondo temperature on the control
parameter $W$ and $\Delta=-\Delta_{\rm ex}$  for odd occupation
(left panel) and even occupation (right panel).} \label{fig3}
\end{figure}

This behavior may be easily understood within a perturbative
(high-temperature) renormalization group (RG) approach. This
approach is based on the study of flow diagrams describing the
evolution of effective coupling parameters $J_{a}(\eta)$ due to
reduction of the effective energy scale $D$ of conduction electron
kinetic energy \cite{And80} (label $a$ enumerates the vertices in
the exchange Hamiltonian, $\eta=\ln D$ is the scaling variable).
The general form of the system of scaling equations is
\begin{equation}\label{rg1}
\frac{dj_a}{d\eta}= -\sum_b c_{ab} j_aj_b
\end{equation}
Here $j_a= \rho J_a$ are dimensionless coupling constants and
$c_{ab}$ are numerical coefficients. These equations should be
solved under the boundary conditions $j(D_0)=\rho J_{a0}$, where
$D_0$ and $J_{a0}$ are the initial conduction bandwidth and the
bare exchange integrals entering the Hamiltonian $H_{\rm ex}$.
When $D\sim D_0 \gg (W,\Delta_{\rm ex})$, one may neglect the
splitting of energy terms in the system (\ref{rg1}),  and all
coupling constants $J_a$ evolve together. When the scale $D$ is
reduced down to the splitting energy $W$ or $\Delta_{\rm ex}$, the
coupling parameters related to the upper level in the multiplets
(\ref{2.1}) or (\ref{2.2}) stop to evolve, because the
corresponding degrees of freedom are quenched below these
energies. These parameters are $K$  both in the Hamiltonians
 (\ref{xeff1}) and (\ref{xeff3}), and the exchange vertices $\tilde J_{ij}$
 related to the singlet-triplet transitions in (\ref{xeff3}). In a degenerate
 model (\ref{xeff2}), $T_K$ is partially suppressed by the fictitious magnetic
 field $(\bf B)$.  As a
 result the temperature $T_K$ obtained as a solution of these
 equations becomes explicit function of the splitting energy. Its
 behavior is illustrated in the left and right panels of Fig.
 \ref{fig3}
for the charge sectors ${\cal N}=1,2$, respectively. In case of
odd occupation ${\cal N}=1$ the point $W=0$ corresponds to two
equivalent dots decoupled from each other. The system remains
quasi degenerate till $W< T_K(0)=D\exp 4/\rho J_0$. With further
increase of the level splitting, the contribution of the state
$E_2$ in Eq. (\ref{2.1}) to the Kondo tunneling diminishes. In the
asymptotic regime $|W|/T_K(0)\gg 1$, the evolution of $T_K$ is
described by the asymptotic equation \cite{Eto00,Eto05}
\begin{equation}
T_K(W)=\frac{T_K(0)^{\gamma +1}}{|W|^\gamma}
\end{equation}
This asymptotic curves are shown bold Fig.3. The exponent depends
on the model parameter and detailed geometry of quantum dot
\cite{Eto00,Eto05,PG01,KKA04}.

Similar asymptotic curve describes also the evolution of $T_K$ for
even occupation ${\cal N}=2$ with $|W|$ substituted for
$\Delta_{\rm ex}$. But in this case such behavior is
characteristic only for $\Delta_{\rm ex}<0$, where the ground
state is spin triplet. $T_K$ is maximal in the "critical point" of
S/T crossover. On the triplet side of S/T transition it
diminishes due to gradual quenching of triplet-singlet excitations
described by the operator ${\bf R}$ in the Hamiltonian
(\ref{xeff3}). On the singlet side of S/T crossover, $\Delta_{\rm
ex}>0$ the Kondo temperature falls down steeply. The
singlet-triplet crossover driven by some external parameter
(magnetic field, gate voltage etc) in DQD with even occupation is
the most salient effect predicted and observed in these systems.

Additional information unveiling  specific features of Kondo
screening in complex quantum dots can be extracted from the
temperature and magnetic field dependence of tunnel conductance
$G_{{\rm max}}$ at given $W$ or $\Delta_{\rm ex}$. Its behavior
was discussed time and again \cite{Bord03,HoSch,PG01,PG04,HoZa}
The most interesting is the behavior of conductance in the charge
sector ${\cal N}=2$.

\underline{On the triplet side} of S/T crossover $G_{\rm max}(T)$
increases with decreasing T. At high $T\gg T_K(|\Delta_{\rm
ex}|)$, it grows logarithmically,
\begin{equation}\label{limh}
G_{\rm max}(T)/G_0\sim \ln^{-2}(T/T_K)
\end{equation}
where $G_0=2e^2/h$ is the  limiting value of tunnel conductance in
a Kondo regime (unitarity limit, where the sum of all phase shifts
on the Fermi levels $\delta_{i\sigma}$ equals $\pi/2$). This
limiting value quantify the tunnel conductance at $T-0$. According
to the Friedel sum rule \cite{Fried,Langr}, generalized for the
two-channel geometry \cite{ISS,PHG} one has
\begin{equation}\label{friedel}
G(T=0)=G_0\sin^2 \left[\frac{\pi}{2}(\bar n_2-\bar n_1 ) \right]
\end{equation}
where $\bar n_{(2,1)}$ is the change in number of electrons under
the Fermi level due to Kondo screening in two channels (even and
odd). In terms of the phase shifts the argument of the sine
function in the r.h.s. of Eq. (\ref{friedel}) reads
$\frac{1}{2}\sum_{i\sigma}\delta_{i\sigma}$. This means that the
demand of spin rotation invariance and \textit{l-r} symmetry means
that the phase shift in each channel reaches $\pi/4$ at $T=0$.
Similarly to the theory of Kondo scattering \cite{Noz74}, the
deviation from the unitarity limit is of standard Fermi-liquid
character
\begin{equation}\label{liml}
G(T)/G_0=\left[1-\pi\left(T/T_K\right)^2 \right]
\end{equation}
at $T\ll T_K$. The monotonous interpolation \cite{PG04} between
two limiting temperature regimes (\ref{limh}) and (\ref{liml})
may be violated due to multistage Kondo effect. We will discuss
this regime, when considering the case of triple quantum dots (see
below).

The influence of external magnetic field $B$ on the tunnel
conductance at low $T\ll B$ is easily tractable \cite{PG04}: in
accordance with the general theory of Kondo effect in presence of
Zeeman splitting of the levels in the dot, in this regime $T$
should be substituted for $B$ in the asymptotic equations
(\ref{limh}) and (\ref{liml}). At low $T\ll T_K$ additional
information may be obtained by means of the NRG method
\cite{HoZa}. Due to the loss of spin rotation symmetry the phase
shifts $\delta_{i\sigma}$ become explicit functions of magnetic
field. In a single channel case this dependence is scaled by
$T_K$, namely $\delta_\sigma (B) = \delta(0)+\sigma (B/T_K)$. In a
two-channel regime, level splitting $W$ between the dot levels
enters this dependence explicitly. A simple equation describing
this dependence was derived analytically \cite{PHG},
\begin{equation}
\delta_{i\sigma}=\delta(0)+\sigma (B/T_i)+(-1)^i(B/T_i^\prime)
+O(B^3).
\end{equation}
Here the parameters $T_i$ and $T_i^\prime$, which scale the field
$B$ depend on $W$. NRG calculations \cite{HoZa} show great variety
of magnitoconductance curves $G(B,W)$.

\underline{On the singlet side} of S/T crossover the ground state
of DQD is nonmagnetic, so that there is no room form Kondo-type
ZBA at $T=0$ (see the right panel of Fig. \ref{fig3}). However, at
$T> \Delta_{\rm ex}$ the triplet state is still involved in
perturbative scaling equations (\ref{rg1}), so that $G(T)$ grows
with decreasing $T$ in accordance with Eq. (\ref{limh}). The
temperature T/S crossover occurs at $T\sim \Delta_{\rm ex}$, and
conductance starts to fall with further decrease of $T$ ending
with exponentially small value of $G(T=0)$ \cite{HoSch}. Similar
effect should be observed in the behavior of $G$ as a function of
source-drain voltage $v_{ds}$. At $ev_{ds}>\Delta_{\rm ex}$ the
DQD shows Kondo tunneling, but with lowering bias the triplet
channel is quenched, and conductance shows up a zero field dip
instead of zero field peak. Apparently this type of crossover was
observed \cite{Wiel} in a two-orbital planar dot with even
occupation.
\subsubsection{T-shape geometry}

In a T-shape geometry one of two dots is detached from the leads
(Fig. \ref{config2}c). In the first experimental device of this
type \cite{Mol95}, the role of the side (right) dot was to control
single-electron tunneling through the left dot. Recently several
new effects related to the Kondo-regime were discovered in this
geometry.

First, it was shown \cite{KA01} that the T-shape double dot with
even occupation ${\cal N}=2$ demonstrates the properties of spin
rotator with $SO(4)$ symmetry. The simplest form of the
Hamiltonian (\ref{xeff3}), namely
\begin{equation}
H_{\rm ex}= 2J{{\bf S}}\cdot {\bf s} + 2\tilde J{\bf R}\cdot {\bf
s}
\end{equation}
was derived just for this model. Besides, it was shown that in the
asymmetric T-shape DQD, where the Coulomb blockade in the right
dot is sufficiently larger than in the left dot, the S/T crossover
may occur because of the many-body logarithmic renormalization
\cite{BKM,Hald}, which is determined by the renormalization group
invariants $E_\Lambda^*$, namely
\begin{equation}\label{barbar}
E_\Lambda^* = E_\Lambda(D) -\pi^{-1}\Gamma_\Lambda \ln(\pi
D/\Gamma_\Lambda)~.
\end{equation}
The level crossing is possible because the inequality $\Gamma_T
> \Gamma_S$ for the tunneling rates $\Gamma_\Lambda$ is realized
for asymmetric DQD, so that the renormalization of  the triplet
level $E_T$ is stronger than that of the singlet state $E_S$.
Similar effect may be achieved in a symmetric DQD by means of the
gate voltage applied to the side dot (later on the possibility of
inducing the S/T crossover by means of the the gate voltage was
found also for two-orbital planar dots \cite{Kogan}).

Another interesting possibility arises at the odd occupation
${\cal N}=1$, where the strong Coulomb blockade exists only in the
side dot \cite{Kang,Kim,Takaz,Torio02}. It was found that in this
case the Kondo resonance arises on the background of otherwise
non-correlated transport between the electrodes via the right dot.
The interference between the resonance scattering and free
propagation is known as Fano effect discovered originally in the
optical absorption spectra of free atoms \cite{Fano}. The Fano
effect in atoms is described by the same Hamiltonian as the
Anderson impurity hybridization \cite{And61}, so there is no
wonder that similar effect was found in the resonance impurity
scattering in metals \cite{Shibat}. Since the single electron
transport through the quantum dot is also described by the
Anderson Hamiltonian (\ref{and2}), the Fano effect ubiquitous in
resonance scattering was expected in tunnel conductance of quantum
dots, and the structures of characteristic Fano-type form were
indeed observed in the tunneling spectra of planar quantum dot
\cite{Gores}. In the latter case the Fano effect arises due to the
interference between the resonance level in the quantum dot and
the band continuum in the leads. The corresponding contribution to
the conductance has the form
\begin{equation}
G_{\rm Fano}(\epsilon)= G_0\frac{(\varepsilon
+q)^2}{\varepsilon^2+1},
\end{equation}
where $\varepsilon=2(\epsilon-\epsilon_d))/\Gamma$,
 $q$ is the so called
asymmetry parameter predetermined by the spectral characteristics
of the lead electrons, $\Gamma=\pi\rho V^2$ is the tunneling rate.
In simple terms, Fano effect is nothing but modification of the
spectral density of conduction electrons due to its repulsion from
the resonance level superimposed on the continuous spectrum.

In a T-shape geometry the Fano effect arises due to superposition
of the {\it Abrikosov-Suhl resonance} created by the Kondo effect
in the right dot on the continuous tunneling spectra of the system
`source -- left dot -- drain'. As was noticed by Kang et al
\cite{Kang}, this Fano-Kondo effect looks as an antiresonance in
conductance: instead of the standard Friedel-Langer formula
(\ref{friedel}), one has
\begin{equation}\label{kang}
G(T=0)=G_0\cos^2 \left[\frac{\pi}{2}\bar n_r \right]
\end{equation}
where $n_r$ stands for the average occupation number of the right
dot. Such "unitarity limit" for conductance means that in the
T-shape geometry, the Kondo effect in the right dot results in
complete suppression of resonance tunneling through the left dot
and  a dip arises in the tunnel conductance instead of the usual
peak. As was mentioned above, the phase shifts in the Kondo regime
are spin dependent, so the modification of tunnel spectra due to
destructive Kondo-Fano interference is also spin-dependent, and
the T-shape dot in this regime may work as a spin filter in
external magnetic field \cite{Torio04}.

One more advantage of the TQD geometry is the possibility to
approach the two-channel Kondo effect \cite{Oreg03,Pustilnik04}.
Such possibility arises in the geometry of Fig. \ref{config2}c,
when the Coulomb blockade is strong in the left dot and the right
dot is big enough so that the level spacing is less than the
tunnel rate, but the Coulomb blockade is still strong and fixes
the electron occupation number. In such conditions the way opens
to overcome the main difficulty in realizing a physical system
that materializes the two channel Kondo model. The necessary
precondition for this regime is in creating two separate channels
that equally screen the spin \cite{Nobla}. In conventional setups
an electron from one channel that hops on the dot may hop to the
other channel and thus cause mixing between the channels. This
mixing lead eventually to two``eigen channels" with one channel
coupled stronger than the other one. The channel with the stronger
coupling fully screen the spin and the other channel is decoupled,
and we thus have again the single channel Kondo case. It was
suggested \cite{Oreg03}  to overcome this mixing problem by using
a large quantum dot as an additional channel.  Then, the free
leads form one channel [even one in accordance with our
classification (\ref{evod})] and the large dot forms the second
channel. The channels do not mix as transfer of electrons between
them charges the large dot. As a result, the SW transformation
maps the original Anderson Hamiltonian for a T-shape quantum dot
onto the two-channel exchange Hamiltonian
\begin{equation}
H_{\rm 2ch}=\sum_{\gamma,k}\epsilon_{\gamma,k}
c^\dag_{\gamma,k}c^{}_{\gamma,k}+\Sigma_\gamma J_\gamma {\bf
S}\cdot {\bf s}_\gamma + BS^z
\end{equation}
Here the channel index $\gamma$ stands for the even lead-dot
channel and the states in the large dot $r$.  The tunnel
conductance in the 2-channel non-Fermi liquid regime is realized
at $J_2=J_r$. It demonstrates specific temperature dependence. In
accordance with predictions of conformal field theory
\cite{Pustilnik04,Affleck93},
\begin{equation}
G_{\rm nfl}(T)= \frac{G_0}{2}\left(1 - \sqrt{\pi
T/T_K}\right)
\end{equation}
Practically, the zero temperature limit cannot be achieved because
the Kondo screening in the large dot is quenched due to
discreteness of its electron spectrum at $T \sim \delta\epsilon$.

Unlike the single channel case,  the magnetic field is a relevant
parameter in the two channel case \cite{Affleck92}. Introducing
the channel anisotropy parameter $\Delta_{\rm ch}=J_r-J_2\ll J_2$,
which describes deviation from the 2-channel fixed point
($\Delta_{\rm ch}=0$), the following equation for the
magnitoconductance may be derived \cite{Pustilnik04,Affleck92}
\begin{equation}
G_{\rm nfl}(T,B)=G_0\left[\frac{1}{2}+a\,{\rm sign}(\Delta_{\rm
ch})\frac{B_\Delta}{B} - b\,\frac{B}{T_K}\ln \frac{T_K}{B}\right]
\end{equation}
where  $B_\Delta = \Delta_{\rm ch}T_K/J_2^2$, and $a,b$ are
positive numerical coefficients of the order of 1
\subsection{Triple quantum dots}

In this section we consider new features of Kondo mapping which
are characteristic for the chains consisting of three dots (Fig.
\ref{config3}a-c). Simple increase of the number of sites in the
chain promises no new qualitative results, so we will discuss here
the configurations where the central dot differs from two side
dots in its size (and hereby by the magnitude of the Coulomb
blockade parameter $Q_c$), but the latter dots are identical, so
that the TQD retains its reflection symmetries. Both conceivable
situations, namely $Q_c\gg Q_s$ and $Q_c\ll Q_s$ will be
considered (the indices $c,s$ are used in this section to indicate
the physical quantities related to the central and side dots,
respectively). In the first experimental realization of TQD
\cite{Marcus} the former option was chosen, whereas the first
theoretical study \cite{KKA04} was devoted to to the second
possibility.

In a TQD with "open" central dot and $Q_c\ll Q_s$, its role in
formation of Kondo tunneling regime reduces to providing the
channel for indirect RKKY-type mechanism of exchange between two
localized spins formed in side dots. Thus, from the theoretical
point of view this problem is in fact may be mapped onto that for
a DQD (see Section III.A) with specific exchange mechanism. The
Kondo effect may be described in terms of two-site Kondo
Hamiltonian, where  the trend to interdot spin coupling competes
with the trend to individual Kondo coupling between two side dots
and the electrons in the leads \cite{Silo,Vava}. One may mention
in this connection the phase diagram of effective two-site Kondo
effect calculated in a framework of the model, where the pair of
spins is coupled to the linear electron chain in a side geometry
\cite{Vojta02}.

Qualitatively new features of Kondo mapping problem were found out
in the opposite limiting case $Q_c\gg Q_s$. As follows from the
general scheme of Kondo-mapping based on the dynamical symmetry of
artificial molecule (Section III.A), the form of effective
exchange Hamiltonian (\ref{xeff3}) depends on the structure of
spin multiplet of isolated multivalley quantum dot. Up to now only
two possibilities were exploited: spin and orbital doublet for odd
occupation ${\cal N}=1$ and single-triplet pair for even
occupation ${\cal N}=2$. Study of TQD with odd and even
occupations ${\cal N}=3$ and ${\cal N}=4$ give us new
opportunities \cite{KKA04,KKA03}.

To demonstrate these opportunities let us consider TQD  in lateral
geometry. Genesis of spin multiplets may be understood from a
general set-up illustrated by Fig. \ref{fig4}.
\begin{figure}[htb]
\centering
\includegraphics[width=75mm,angle=0]{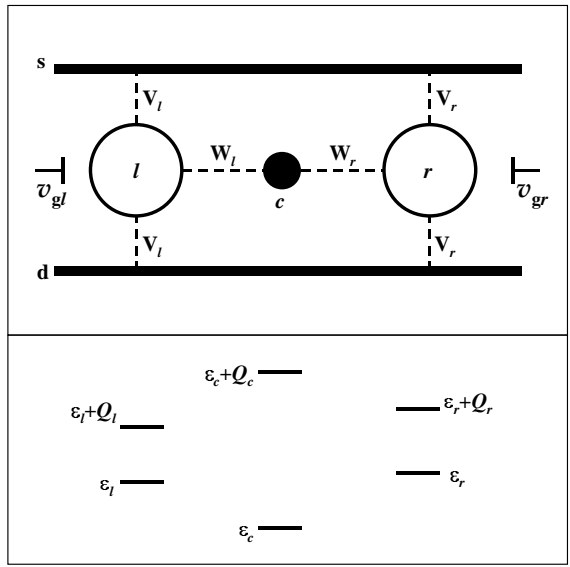}
\caption{TQD in parallel geometry and energy levels of each dot
$\varepsilon_a-ev_{ga}$ at $W_a=0$.} \label{fig4}
\end{figure}
In case of ${\cal N}=3$ three electrons are distributed over three
dots in such a way that the state with doubly occupied central dot
is suppressed by strong Coulomb blockade $Q_c$. In accordance with
the Young tableaux for a system with \textit{l-r} permutation
symmetry, the spin multiplet consists of two doublets with $S=1/2$
having even and odd symmetry relative to this permutation and one
quartet $S=3/2$ with full orbital symmetry. In case of ${\cal
N}=4$ the spin manifold consists of two spin singlets $S_{e,o}$
and two triplets $T_{e,o}$, both even and odd relative to
\textit{l-r} permutation. Varying the gate voltages
$v_{gl},v_{gr}$ and playing with tunnel parameters $V_{l,r}$ and
$W_{l,r}$, one may break \textit{l-r} symmetry (Fig. \ref{fig4})
and change the singlet-triplet splitting, so that the spin states
are classified as $S_{l,r}$ and $T_{l,r}$. The relative positions
of energy levels in spin multiplets evolve as a function of model
parameters and various types of level crossings occur (see
\cite{KKA04} for detailed calculations). Similar situation arises
for vertical configuration of Fig. \ref{config3}a.

 In accordance with general theory of dynamical symmetries \cite{Nova},
quasi degeneracy of low-lying states in spin multiplets within the
energy scale $\sim T_K$  generates special symmetries of TQD. For
example, if the multiplet of low-lying states consists of two
singlets and one triplet, the relevant dynamical symmetry is
$SO(5)$. If this multiplet is formed by two triplets and one
singlet, the corresponding symmetry is $SO(7)$, etc. The methods
of constructing the generators for these groups are described in
details in the reviews \cite{Nova,kisrew}. As a result, an unique
opportunity arises to change the value of index $n$ characterizing
the symmetry $SO(n)$ of TQD by varying the gate voltages and other
experimentally controllable parameters of a device. The phase
diagram of vertical TQD with ${\cal N}=4$ calculated in
\cite{KKA04} is presented in Fig. \ref{fig5}.
\begin{figure}[htb]
\centering
\includegraphics[width=75mm,angle=0]{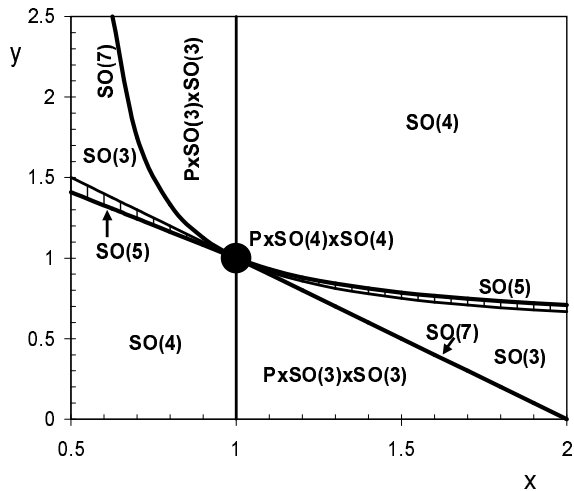}
\caption{Dynamic symmetries in TQD: Phase diagram in coordinates
$x=\Gamma_l/\Gamma_r$,
$y=(\varepsilon_l-\varepsilon_c)/(\varepsilon_r-\varepsilon_c)$.}
\label{fig5}
\end{figure}
This diagram shows great variety of phases with different
symmetries from the most symmetric one $P_2\otimes SO(4)\otimes
SO(4)$ to conventional $SO(3)$ phase where the ground state of TQD
is spin triplet, or non-Kondo  singlet ground state (shaded
areas). Each phase is characterized by its own $T_K$, and this
means that the ZBA in conductance should follow the change of the
Kondo temperature, so that each crossover from one symmetry to
another is accompanied by the abrupt change of conductance at
given temperature.

At ${\cal N}=3$ we meet a somewhat unexpected situation where
Kondo tunneling in a quantum dot with {\it odd} occupation
demonstrates the exchange Hamiltonian of a quantum dot with {\it
even} occupation. The reason for this scenario is the specific
structure of the wave function of TQD with $N=3$. The
corresponding wave functions are vector sums of states composed of
a "passive" electron sitting in the central dot and
singlet/triplet (S/T) two-electron states in the $l,r$ dots. Then
using certain Young tableaux \cite{KKA04}, one concludes that the
spin dynamics of such TQD is represented by the spin 1 operator
${\bf S}$ corresponding to the $l-r$ triplet, the corresponding
R-operator ${\bf R}$ and the spin 1/2 operator ${\bf s}_c$ of a
passive electron in the central well. The latter does not enter
the effective Hamiltonian $H_{\rm ex}$ but influences the
kinematic constraint via Casimir operator ${\cal C}={\bf
S}^{2}+{\bf M}^{2}+{\bf s}^{2}_c=\frac {15}{4}$. The dynamical
symmetry is therefore $SO(4) \otimes SU(2)$, and only the $SO(4)$
subgroup is involved in Kondo tunneling. Similar situation,
although for different reasons is realized in fork and cross
geometries (see below).

Remarkable symmetry reduction occurs in external magnetic field
\cite{KKA04}. First example of such reduction was found in a
situation where the exchange splitting of S/T multiplet (symmetry
group $SO(4)$) is compensated by the Zeeman splitting \cite{PAK},
so that the up spin projection $|T1\rangle$ of triplet forms a
pseudospin with singlet $|S\rangle$ and the symmetry reduction
$SO(4)\to SU(2)$ takes place. In case of TQD with $SO(5)$
symmetry, due to the same compensation the system may be left in a
subspace $\{T1_l,S_l,S_r\}$. The symmetry reduction in this case
is $SO(5)\to SU(3)$, and the Anderson Hamiltonian is mapped on a
very specific \textit{anisotropic} Kondo Hamiltonian involving
only operators ${\bf R}_i$,
\begin{equation}\label{su3}
H_{\rm ex}= \sum_{ij}\sum_{\mu\nu}J_{ij}^{\mu\nu}R_i^\mu s_j^\nu
\end{equation}
where $\mu,\nu$ are cartesian components of scalar product. Here
the Kondo effect is described exclusively in terms of dynamical
symmetry.

Another non-standard manifestation of Kondo mapping for linear TQD
is the possibility of two-channel Kondo effect in vertical
geometry of Fig. \ref{config3}a with $Q_c\gg Q_s$ at ${\cal N}=3$
with preserved \textit{s-d} mirror symmetry \cite{KKA03}. The
strong Coulomb blockade in central dot prevents direct
\textit{s-d} tunneling. The cotunneling is possible only because
the wave functions of electrons centered on the side dots have
small tails on the central dot. It is crucially important that the
standard rotation (\ref{evod}) does not eliminate the odd channel
from the tunneling Hamiltonian in TQD. In the situation, where the
ground state of TQD is the spin doublet with even parity
$|D_e\rangle$, the SW transformation for the original Anderson
Hamiltonian results in anisotropic two-channel exchange
Hamiltonian,
\begin{equation}
H_{\rm ex}=J_s {\bf S}\cdot {{\bf s}_s}+J_d {\bf S}\cdot {{\bf
s}_d}+J_{sd} {\bf S}\cdot ({\bf s}_{sd}+{\bf s}_{ds})
\end{equation}
Due to the presence of nondiagonal vertex $J_{sd}$ the incurable
orbital anisotropy arises: the tunneling through two channels is
controlled by the parameters $J_{\pm}=(J_s+J_d)/2\pm
\sqrt{(J_s-J_d)^2/4+J_{sd}^2)}$. In accordance with the theory of
two-channel Kondo effect \cite{Nobla}, this anisotropy makes the
2-channel fixed point unattainable, but due to strong Coulomb
blockade in central dot $J_{sd}=\sim V^2W^2/Q_c\varepsilon_d^2$ is
extremely small, so that one may approach the fixed point close
enough, and the predecessor of 2-channel regime may be observed
experimentally as a dip in conductance $G$ as a function of the
difference of gate voltages $v_{gs}-v_{gd}$ applied to the side
dots (this difference controls the degree of channel anisotropy).

\subsubsection{TQD in cross and fork geometry}

To complete the studies of linear artificial molecules we consider
in this section the configurations shown in Fig.
\ref{config3}(c,f). We have seen above that one may meet the
situation, where   the linear TQD with odd occupation and
half-integer spin demonstrates the Kondo physics characteristic
for even occupation with integer spin due to the fact that one of
the electrons in the dot does not participate in tunneling. Here
we will discuss two more mechanisms of such "disguise"
\cite{KKA06b}.

One of these mechanisms is realized in \underline{cross geometry}
at occupation ${\cal N}=3$ under condition $Q_s \gg Q_d$ for
Coulomb blockade parameters. In this case two side electrons are
passive: the tunneling between source and drain occurs through the
central dot. However, these passive electrons influence Kondo
mapping because they are responsible for the \textit{parity} of
the 3-electron wave function relative to the \textit{l-r}
reflection. Diagonalization of the low-energy spin states shows
that it consists of three spin doublets and one spin quartet, and
the lowest state in this manifold is the doublet $D_u$, which is
odd relative to the mirror reflection (see \cite{KKA06b} for
details). Although the wave functions of two passive electrons do
not enter explicitly in the indirect exchange integral arising due
to SW transformation, this integral changes its sign due to odd
parity of the state $D_u$. Thus, in contrast to the standard
paradigm of Kondo mapping, the effective exchange Hamiltonian
corresponds to ferromagnetic coupling, which is irrelevant to
Kondo effect, and the Kondo-type ZBA does not arise in this case
in spite of the fact that the net spin of quantum dot is 1/2.

However, this is not the end of the story. The excited states in
the spin multiplet which are Kondo active, influence the tunnel
transparency and conductance at finite temperature and finite
energies of incident electron due to non-trivial dynamical
symmetry of TQD described above. The states involved in the Kondo
effect are the even spin doublet $D_g$ and the quartet $Q$, so
that the overall dynamical symmetry of  TQD in cross geometry is
$SU(2)\otimes SU(2)\otimes SU(2)$. We meet here the situation,
which reminds the "two-stage" Kondo effect in DQD with ${\cal
N}=2$, on the singlet side of S/T crossover (see Fig. \ref{fig3}b
and subsequent discussion). Here, however, there are three stages
of Kondo screening, where the states $Q$ and $D_g$ are quenched
one after another with decreasing energy or temperature and the
Kondo screening eventually stops at zero $T$. Besides, the
hierarchy of tunneling rates $\Gamma_Q > \Gamma_{Dg}> \Gamma_{Du}$
exists in this charge sector, so the level crossing controlled by
the parameters of TQD is possible in accordance with Eq.
(\ref{barbar}). This level crossing is shown in Fig. \ref{fig6}.
Here the scaling variable is chosen in the form $\eta=\ln (\pi
D/\Gamma_Q)$. The value of $\bar D$ is determined from the
crossover condition $\bar D_\Lambda \approx E_\Lambda(\bar
D_\Lambda)$, where the renormalization (\ref{barbar}) changes for
the SW regime with fixed charge, $\bar D_0$ is the initial value
of scaling variable.
\begin{figure}[htb]
\centering
\includegraphics[width=70mm,height=50mm,angle=0]{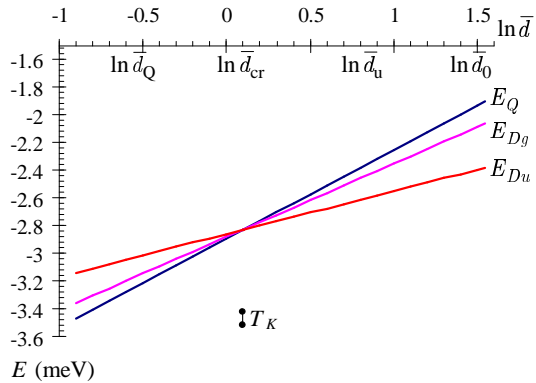}
\caption{Flow diagram for the levels $E_{\Lambda}$  determined by
the scaling invariant (\ref{barbar}). $ \Lambda=D_u, D_g, Q$,
$\bar d=\pi \bar D/{\Gamma_Q}$. Energy is measured in meV units
(see text for further explanations).} \label{fig6}
\end{figure}

Three points ($\ln \bar D_u,~ \ln \bar D_Q,~ \ln \bar D_{cr},$) on
the abscissa axis correspond to three values of the control
parameters where the crossover to the SW regime occurs for the
ground states $E_{D_u}$, $E_Q$ and the completely degenerate
ground state, respectively. By means of appropriate variation of
the control parameters, the system may be transformed from a
non-Kondo regime with the ground state $E_{Du}$ to the
underscreened Kondo regime with the ground state $E_Q$ and spin
$S=3/2$. In accordance with the general theory of Kondo mapping
(Section III.A), $T_K$ is maximum in the point of maximum
degeneracy. Evolution of $T_K $ is shown in Fig. \ref{fig7}, which
should be compared with Fig. \ref{fig3}b.
\begin{figure}[htb]
\centering
\includegraphics[width=70mm,height=50mm,angle=0]{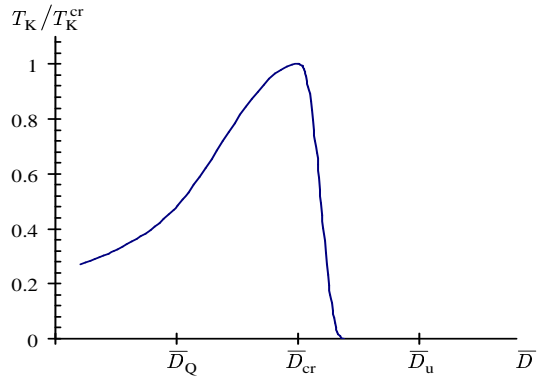}
\caption{Evolution of $T_K$ as a function of a control parameter
$\bar D$.} \label{fig7}
\end{figure}
Unlike the case of singlet/triplet crossover in DQD with ${\cal
N}=2$, here one deals with the crossover from the non-Kondo spin
doublet $D_u$ to the Kondo spin quartet $Q$ via the highly
degenerate region of $SU(2)\otimes SU(2)\otimes SU(2)$ symmetry.

Let us turn to the \underline{fork geometry} shown in Fig.
\ref{config3}f. In this geometry non-trivial Kondo physics arises
already in the simplest case of odd occupation ${\cal N}=1$ in a
situation with the \textit{l-r} mirror symmetry. The fork may be
considered as a "quantum pendulum" \cite{Sarag,Kradl}. Three
one-electron eigenvalues are
\begin{equation}
E_{D_{b,a}}=\epsilon_c \mp 2W^2/\Delta, ~~~ E_{D_n} = \epsilon_s~,
\label{sps}
\end{equation}
($\Delta=\max\{|\epsilon_s-\epsilon_c|,|Q_c-Q_s|\}$) The
eigenstates are classified as a non-bonding spin doublet $D_n$
(odd combination of the wave functions centered in the sites 1,2)
and bonding/antibonding pair $D_{b,a}$ of corresponding even
combination with the state centered in the cite 3 (see
\cite{KKA06b} for details). The latter pair is the analog of
resonant valence bonds (RVB) known in "natural" molecules. To
describe this pendulum one should introduce the pseudospin vector
$\bf T$ defined in Eq. (\ref{pseudo})  and work with the
Hamiltonian (\ref{xeff1}). Like in the cross geometry, the level
crossing effect as a function of control parameter takes place
(Fig. \ref{fig8}).
\begin{figure}[htb]
\centering
\includegraphics[width=70mm,height=50mm,angle=0]{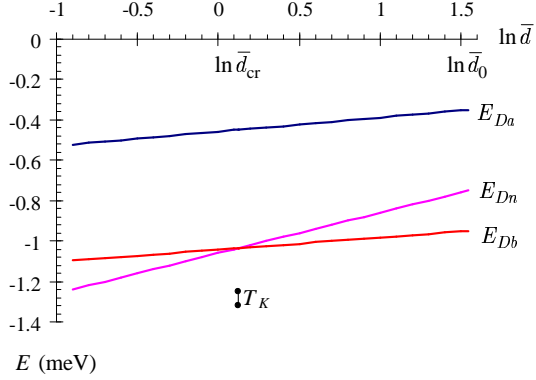}
\caption{Flow diagram for the levels $E_\Lambda$ of the TQD in
fork geometry.} \label{fig8}
\end{figure}

Here $T_K$ is nonzero on both sides of the crossover and its
evolution is described by the bell-like curve similar to that in
Fig. \ref{fig3}a (although slightly asymmetric). However the
tunnel conductance is drastically influenced by the pendulum
structure of the electron wave function. In the three-terminal
fork geometry, one should consider separately the situations,
where the bias voltage is applied between the leads 1 and 2 and
between the leads 1 and 3. We define the corresponding components
of tunnel conductance as $G_{22}$ and $G_{33}$, respectively. The
Kondo  anomaly in $G_{22}$ is predetermined by the RVB pair, and
the ZBA roughly follows the evolution of $T_K$ through the
crossover. More peculiar behavior is expected in 1-3 channel
because the non-bonding state $D_{n\sigma}=
2^{-1/2}(d^\dag_{1\sigma}-d^\dag_{2\sigma})$ is detached from the
lead 3. As a result the Kondo contribution to $G_{33}$ manifests
itself as a finite bias anomaly (FBA) in a situation where the
ground state of TQD is $E_{D_n}$. Tunnel conductance as a function
of bias voltage in both channels is illustrated in Fig.
\ref{fig9}. The dip in the curve \textit{b} on the right panel
reminds similar dip in the tunnel conductance of DQD with ${\cal
N} =2$ on the singlet side of S/T crossover \cite{HoSch,Wiel}.
\begin{figure}[htb]
\centering
\includegraphics[width=65mm,height=35mm,angle=0]{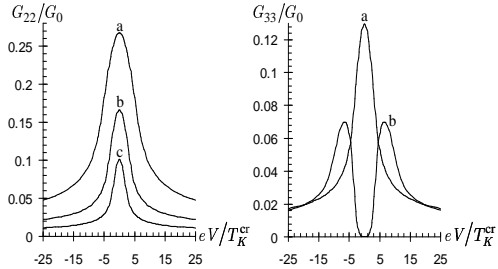}
\caption{Left panel: Tunnel conductance in the channel '2--1' for
$\bar D=\bar D_{cr}$ (a), $\bar D>\bar D_{cr}$ (b) and $\bar
D<\bar D_{cr}$ (c). Right panel: Tunnel conductance in the channel
'3--1' for $\bar D>\bar D_{cr}$ (a) and $\bar D<\bar D_{cr}$ (b).}
\label{fig9}
\end{figure}

Thus, we see that some manifestations of the Kondo effect in TQD
with odd occupation may mimic those for DQD with even occupation
due to specific influence of the mirror reflection on the
structure of the electron wave functions in trimers.

\section{Kondo physics for small rings}

In previous sections we discussed the Kondo effect in short
"Hubbard chains" in contact with metallic reservoirs. Meanwhile,
the Hubbard-like objects were studied also in closed ring
geometries as well (see, e.g., \cite{Hubring1,Hubring2}).
Experimentally Kondo effect on closed rings was observed on
gutter-like dots in planar geometry \cite{Enss04,Enss05}, but the
impact to theoretical investigation of Kondo effect in ring-like
nanoobjects \cite{Brat} was given by experimental studies of Co
trimers adsorbed on metallic sublayer \cite{Jam}. It was shown
that the basic symmetry of equilateral triangular triple quantum
dot (TTQD, Fig. \ref{config3}e) with odd occupation is $SU(4)$ due
to the interplay between the spin and orbital degrees of freedom,
similar to that in two-orbital DQD (see Section III.A). Special
attention was paid to the case  ${\cal N}=3$ which models
triangular Co trimer \cite{Affleck,Savka}. In this charge sector
the effective spin Hamiltonian contains not only exchange
interaction between the spins in the dots and adjacent leads, but
also the two-site Heisenberg exchange between spins in the
neighboring dots. Magnetic frustrations in triangular geometry
affect the spin state and therefore influences the Kondo-type ZBA
in tunneling spectra. These spectra were calculated by the
numerical RG and quantum Monte-Carlo methods. Besides, it was
found \cite{Affleck} that in case of complete channel isotropy the
non-Fermi-liquid regime arises from the interplay of magnetic
frustrations and Kondo physics.

Another phenomenon, which interplays with the Kondo physics is the
Aharonov-Bohm oscillation of tunnel transparency in magnetic field
directed perpendicularly to the plane of triangle. This effect may
 be seen already for the Hubbard ring with 1/3 occupation (TTQD with
${\cal N}=1$), where there is no room for exchange interaction
between spins localized in neighboring sites and concomitant
magnetic frustrations. The starting point for solving the problem
of interplay between Kondo and Aharonov-Bohm phenomena
\cite{KKA06a,KKA05} is the Anderson Hamiltonian (\ref{and2})
rather than the exchange Hamiltonian (\ref{xeff1}).

In accordance with the general scheme discussed in Section II, one
should start with the diagonalization of the Hamiltonian of 3-site
Hubbard ring. The point symmetry of this equilateral triangle is
$C_{3v}$. This group describes the "orbital" degrees of freedom,
whereas the continuous spin symmetry is usual $SU(2)$ symmetry of
spin 1/2. According to the irreducible representations of $C_{3v}$
group, the spectrum of TTQD consists of three levels $\Lambda=DA,
DE_{\pm}$. Here as usual $D$ stands for spin doublet, $A$ is the
fully symmetric orbital singlet and $E_{\pm}$ are two components
of orbital doublet. The energies of these states in out-of-plane
magnetic field $B$ are
\begin{equation}\label{emag}
E_{D\Gamma}(p)=\epsilon-2W\cos \left(p-\frac{\Phi}{3}\right).
\end{equation}
such that for negative $W$ and for $B=0$, $p=0,~2\pi/3,~4\pi/3$
correspond respectively to $\Lambda=A,E_\pm$ with the ground state
$DA$, so that the orbital degrees of freedom are quenched at low
temperature and the SW mapping procedure ends with conventional
Kondo Hamiltonian (\ref{hex}).  However, using the magnetic field
as a control parameter, one may initiate level crossing by varying
the magnetic flux $\Phi$ through TTQD. This  level crossing is
shown on the upper panel of Fig. \ref{fig10}.
\begin{figure}[htb]
\centering
\includegraphics[width=50mm,height=50mm,angle=0]{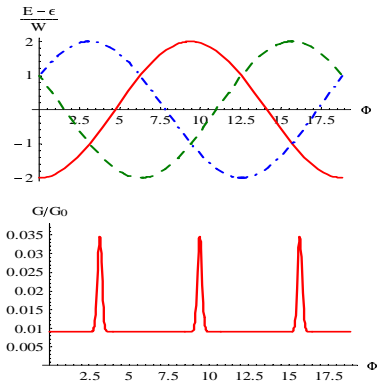}
\caption{Upper panel: Evolution of the energy levels $E_A$ (solid
line) and $E_\pm$ (dashed and dash-dotted line, resp.) Lower
panel: corresponding evolution of conductance ($G_0=\pi e^2/\hbar
$).}\label{fig10}
\end{figure}
In each crossing points and its nearest vicinity the dynamical
symmetry of TTQD is SU(4), so that the symmetry crossovers
$SU(2)\to SU(4) \to SU(2)$ occur at $\Phi=(n+\frac{1}{2})\Phi_0$,
where $\Phi_0$ is the quantum of magnetic flux. Each crossover is
accompanied by the change of $T_K$ from $\exp(-1/2J)$ to
$\exp(-1/4J)$ and back (see Section III.A). The ZBA peak in the
two-terminal tunnel conductance changes accordingly (Fig.
\ref{fig10}, lower panel).

Although formally there are three tunneling channels, the
non-Fermi liquid Kondo regime cannot arise, because the
non-diagonal components $J_{ij}$ appear in the exchange
Hamiltonian (\ref{xeff1}). Further diagonalization should be done
by means of rotating frameworks for Bloch electrons. This
diagonalization introduces irremovable channel anisotropy, so that
the 2-channel non-Fermi-liquid fixed point cannot be achieved
unlike the case considered in \cite{Affleck}, where the channel
isotropy was postulated from the very beginning.

In order to realize the Aharonov-Bohm interference, one should use
the two-terminal geometry shown in Fig. \ref{config3}d. In this
case there are two paths (1-3) and (2-3) for single electron
tunneling between source and drain. Interference of two waves in
the drain results in famous Aharonov-Bohm oscillations. The field
$B$ affects the lead-dot hopping phases. In the chosen gauge, the
hopping integrals are modified as, $W\to W\exp (i
\Phi_1/3),~~V_{1,2}\to V_s\exp[\pm i (\Phi_1/6+\Phi_2/2)],$ where
$\Phi_{1,2}$ are magnetic fluxes through the upper and lower loop
of the device. As a result the exchange Hamiltonian reads
\begin{eqnarray}
H&=&J_s {\bf S}\cdot {\bf s}_s +J_d {\bf S}\cdot {\bf
s}_d%\nonumber\\
+J_{sd}{\bf S}\cdot ({\bf s}_{sd} + {\bf s}_{ds}) + K{\bf T}\cdot
{\bf t} \label{HAB}
\end{eqnarray}
(the latter term becomes actual when the magnetic field induces
level crossing in accordance with Fig. \ref{fig10}). Magnetic flux
enters the coupling constants $J_s$, $J_d$, $J_{sd}$, $K$ via SW
transformation. As a result the constant $J_{sd}(\Phi_1,\Phi_2)$
turns into zero at some values of magnetic flux, so that the
Aharonov-Bohm interference completely blocks Kondo transparency.
Thus TTQD serves simultaneously as a Kondo "pass valve" and as an
Aharonov-Bohm interferometer. It should be stressed that both the
continuous spin degrees of freedom and discrete "rotations" of
triangle are involved in  these two phenomena in TTQD.
%\section{Charge-spin conversion due to dynamical symmetries}
\section{Concluding remarks}

Among many aspects of Kondo tunneling through complex quantum dots
we have chosen for this review only the symmetry related
properties predetermined by the structure of the low-lying states
in the spin multiplet characterizing the fixed charge sector of
complex quantum dot. New features, which are introduced by the
dynamical symmetries in the Kondo physics are the multistage
process of Kondo screening, symmetry crossovers driven by
experimentally tunable control parameters, interplay between
continuous spin rotation symmetry and discrete point symmetry of
nanodevices. The main tool of experimental monitoring of variable
symmetries is study of temperature and magnetic filed dependence
of zero- and finite-bias anomalies in tunnel conductance.

Among other facets  of Kondo effect in nanostructures one should
mention non-equilibrium Kondo effect at finite bias and under
light illumination, where both quantum dots and leads are far
enough from thermodynamic balance. Under this conditions such
phenomena as spin relaxation, dephasing and decoherence influence
the tunnel transport in Kondo regime. Real atoms and molecules
also may be included in electric circuit by means of advanced
experimental techniques (scanning tunnel spectroscopy,
break-junction method etc). In this case phonon- and photon
assisted processes should be taken into account, which result in
interplay of Kondo resonance tunneling with various "polaronic"
and "excitonic" effects. Besides, Kondo processes may be included
in nano-electro-mechanical shuttling, so that Kondo physics of
movable objects becomes one of challenging items on the agenda.

\end{document}